\documentclass{article}

\usepackage[position, preprint]{neurips_2026}

\usepackage[utf8]{inputenc} % allow utf-8 input
\usepackage[T1]{fontenc}    % use 8-bit T1 fonts
\usepackage{hyperref}       % hyperlinks
\usepackage{url}            % simple URL typesetting
\usepackage{booktabs}       % professional-quality tables
\usepackage{amsfonts}       % blackboard math symbols
\usepackage{nicefrac}       % compact symbols for 1/2, etc.
\usepackage{microtype}      % microtypography
\usepackage{xcolor}         % colors
\usepackage{tikz}

\title{Beyond Cash Flows: A Multi-Agent AI Framework for Valuing Clinical-Stage, Cross-Border Biotechnology}

\author{%
  Yuhan Fang \\
  Strategic Investment Manager \\
  CPC Scientific Inc. \\
  \texttt{} \\
}

\begin{document}

\maketitle

\begin{abstract}
  A new class of software systems is transforming investment analysis. Large language model agents assembled into collaborative team structures including analysts, researchers, and risk managers are increasingly deployed across financial markets. Yet current multi-agent frameworks share a critical limitation: they rely on the foundational assumption that companies can be valued through traditional cash flows. This paradigm fails in clinical-stage biotechnology, where enterprise value depends entirely on binary scientific and regulatory milestones. To bridge this gap, this paper introduces a specialized multi-agent framework. Its valuation layer translates qualitative scientific judgment into defensible valuations for pre-revenue assets; its cross-market coordination layer reconciles pricing across international venues simultaneously; and its conflict-fusion mechanism systematically arbitrates between bullish scientific conviction and cautious regulatory constraints in a domain-specific manner. Crucially, the architecture is not a speculative design: it encodes a method the author first executed by hand as sole portfolio manager of China’s first dedicated cross-border biotechnology fund, a human practice that returned 127.17\% against a 50.67\% benchmark within sixteen months. That record is evidence for the underlying method rather than for any AI system; no implementation is evaluated here. This paper presents the framework at the architectural level, establishing foundational design principles for extending agentic investment systems into complex, event-driven asset classes they currently serve poorly.
\end{abstract}

\section{Background}

A new class of software systems now automates the work of an investment team. Large language model (LLM) "agents" are assembled into structures that mirror a fund \citep{Xiao2024, virattt2024, Zhao2025, Zhou2024}: analyst agents gather information, researcher agents debate, a risk agent constrains exposure, and a portfolio-manager agent renders a final decision. Building on general advances in LLM-based multi-agent collaboration \citep{Park2023, Wu2024, Hong2024, Guo2024} and on domain-adapted financial language models \citep{Wu2023, Yang2023}, these systems are real, increasingly capable, and have the potential to replace human teams.

However, it breaks on the assets that matter most to biomedical innovation. Every leading multi-agent investment system shares a hidden assumption: that a company can be valued from its earnings and cash flows. Their valuation logic runs on discounted cash flow (DCF), owner earnings, price-to-earnings ratios, or EV/EBITDA multiples \citep{Damodaran2012}. All of which require a company to have revenue or profit. Clinical-stage biotechnology companies have neither. A pre-revenue biotechnology firm's entire value rests on binary scientific and regulatory events: will a drug pass its Phase III trial? will a regulator approve it? Such events are neither rare nor marginal: fewer than 14\% of drugs entering Phase I are ultimately approved \citep{Wong2019, Hay2014}, and roughly half of new molecular entities fail to secure approval on first submission to the regulator \citep{Sacks2014}. Point a conventional valuation agent at such a company and it returns an unreliable number that is precise-looking but resting on inputs its method cannot supply.

Therefore, this paper proposes a multi-agent framework built specifically for that gap. Its contribution is not the team-of-agents metaphor, which is now well established, but three things that established systems do not do: 
\begin{itemize}
    \item A valuation layer that translates scientific and clinical judgment such as mechanism of action, trial data, regulatory path into a defensible value for a company with no earnings;
    \item A cross-market coordination layer that reconciles valuation, liquidity, currency, and conflicting regulation across the A-share, Hong Kong, and U.S. markets simultaneously, where persistent price divergences between economically identical claims are a well-documented feature rather than an anomaly \citep{Froot1999, Mei2009, Karolyi2006};
    \item A domain-specific mechanism for fusing conflicting expert opinions, e.g., how to weigh a bullish scientific read against a cautious regulatory read, which goes beyond the generic voting and debate used by existing systems \citep{Du2024}.
\end{itemize}

Crucially, this framework is not a paper design in search of validation. The judgment it seeks to encode is one the author first exercised by hand. Between 2019 and 2021, as sole portfolio manager of the first cross-border biotechnology fund in the People's Republic of China (CSRC fund code 001984), the author applied a repeatable, three-dimensional method (set out in Section~\ref{sec:glocal} as the "Glocal" practice) for pricing exactly this asset class, ahead of the market's own infrastructure for doing so. That human practice, and its documented results, are the empirical foundation on which this paper rests: the multi-agent architecture that follows is the author's attempt to scale, rather than to invent, a method already proven to work.

The paper describes this framework at the level of architecture and method. \textit{It deliberately does not disclose the proprietary parameters, weights, or implementation details that would constitute a tradeable model; those remain subject to professional confidentiality.} What follows is a blueprint for how agentic investment systems can be extended to an asset class they currently serve poorly, and an argument that doing so requires a combination of scientific, cross-market, and investment expertise that explains why the gap has remained open.

\section{The Problem and Opportunity}

The hardest and most consequential input to a biotechnology valuation is a \textit{scientific} quantity such as the probability that a specific drug will work and be approved. It must be produced by reasoning over clinical data, mechanism, and regulatory precedent, not by extrapolating financials. This is the parameter that the established risk-adjusted valuation literature identifies as dominating the result \citep{Stewart2001, Villiger2005}, and around which industry-wide base rates vary by more than an order of magnitude across indications and phases \citep{Wong2019, Hay2014, DiMasi2016}. A valuation system that cannot generate that input can still emit a number, but not a reliable one. Its output will only ever be as good as a scientific judgment it did not actually make. It is not a matter of using a different formula; it is a matter of possessing a different kind of judgment.

The human process for making these judgments is slow and does not scale. A portfolio manager (PM) synthesizes inputs from many sources: sell-side and buy-side analysts, key opinion leaders among clinicians, regulatory specialists, primary scientific literature, conference readouts, and market data. She reads reports, interrogates experts, reconciles disagreements, and integrates all of it into a position. This is high-skill work, but it is bounded by human bandwidth that a PM can cover only so many companies, read only so many papers, and hold only so many conversations. It is also subject to well-documented cognitive biases: anchoring and recency \citep{Tversky1974}, confirmation \citep{Nickerson1998}, and overconfidence in the resulting positions \citep{Barber2001}.

The appeal of a multi-agent system is that it promises to scale and de-bias parts of this process: to read more, cover more, and surface disagreement explicitly. That appeal is precisely why such systems have proliferated for conventional equities. The difficulty is that the biotechnology version of the PM's job depends on the one capability those systems lack: the ability to price value that does not yet exist as cash. Before automating that job, it is worth establishing that the job can be done at all that the judgment is real, repeatable, and produces results. 

\section{Related Work and Challenges}

The idea of assembling LLM agents into a structure mirroring an investment team has become one of the most active areas in applied AI for finance. It inherits directly from the broader literature on LLM-based multi-agent collaboration, in which role specialization, conversational orchestration, and structured debate have been shown to outperform monolithic prompting on complex tasks \citep{Park2023, Wu2024, Hong2024, Du2024, Guo2024}. Several notable systems have advanced this paradigm in the financial setting: \textbf{TradingAgents} \citep{Xiao2024} simulates a full trading firm, organizing seven specialized roles ranging from diverse analysts and multi-round bull-bear researcher debates to a multi-perspective risk team—under a fund manager, offering a mature model for agent conflict resolution. Similarly, \textbf{ai-hedge-fund} \citep{virattt2024}, a prominent open-source project, mimics a hedge fund decision hierarchy where valuation agents compute quantitative intrinsic values (e.g., DCF, Buffett-style owner earnings) alongside sentiment and technical agents, culminating in a risk-managed portfolio decision. Other frameworks build upon similar collaborative structures: \textbf{AlphaAgents} \citep{Zhao2025} employs structured debate and risk-tolerance modeling for equity portfolio construction, while \textbf{FinRobot} \citep{Zhou2024} provides an open-source framework for equity research and valuation via specialized agents. Related work has explored layered memory and persona design for trading agents \citep{Yu2024, Li2023}, built atop finance-specific foundation models \citep{Wu2023, Yang2023}. These systems have collectively worked out much of the hard architecture: how to specialize agent roles, how to structure their collaboration, and how to resolve disagreements through debate, facilitation, or managerial arbitration. \textbf{This paper builds upon this foundation rather than seeking to reinvent it.}

A separate body of work applies AI to the \textit{scientific} questions underlying biotechnology valuation, extending a longer line of machine learning research in drug discovery and development \citep{Vamathevan2019, Jumper2021}. Systems such as AutoCT \citep{Liu2025} and Tx-LLM \citep{Zambrano2024} predict clinical-trial outcomes and therapeutic properties, while commercial platforms like Intelligencia AI \citep{Intelligencia2024} assess the probability of technical and regulatory success across drug-development programs to inform investment decisions. In effect, these systems estimate the probability that a given scientific program will succeed. However, these two bodies of work remain disconnected. Existing multi-agent investment systems feature sophisticated team architectures, yet their valuation layers rely entirely on traditional earnings and cash flows—rendering them ineffective or silent for pre-revenue assets. Conversely, clinical-AI systems model scientific probabilities but lack investment-decision logic: they neither construct portfolios, coordinate across cross-border markets, nor synthesize conflicting expert opinions into capital allocations. To the our best knowledge, no published system occupies the intersection: a multi-agent investment framework with a valuation layer tailored for binary, event-driven, pre-revenue assets, coordinated across multiple national markets. 

\section{Method}

\subsection{Underlying Method: The "Glocal" Practice and Its Track Record}
\label{sec:glocal}

We first give a repeatable, three-dimensional method named "Glocal" with explicit inputs, decision logic, and outputs for recognizing investment opportunities as it appeals and positioning capital inside it, ahead of the market's own recognition of the opportunity. This method was publicly used in a June 2020 interview centered on Yuhan Fang's investment philosophy. The original contribution is its operationalization as an investment framework for cross-border, pre-revenue biotechnology. In a summary: for any candidate company or asset class, the method takes a defined set of inputs (primary-source founder history, the global reach of a company's operating resources, cross-border regulatory and development-velocity data) and applies a fixed decision logic to produce specific, checkable portfolio-construction outputs: a re-priced competitive standing, an adjusted valuation and position size, and a liquidity constraint that makes the resulting conviction position executable without vehicle-level risk. Details are listed below:

\textbf{Dimension 1: }Global Resource and Founder-Capability Repricing. \textit{Input:} the founder's and leadership team's complete professional history, drawn from primary disclosure documents (IPO prospectuses, founder letters) and, where available, a multi-year record of public statements; and a direct assessment of whether the company's operating resources such as human capital, clinical-trial site network, and scientific/clinical data assets. \textit{Decision logic:} score the founding team on strategic breadth, depth of demonstrated execution capability, and durability of strategic orientation over time; independently, determine which peer set the company should be priced against. \textit{Output:} a composite founder-capability score and, where resource globality is confirmed, an upward adjustment to the company's competitive-positioning multiple relative to a domestic-only comparison.

\textbf{Dimension 2:} Cross-Border Regulatory and Clinical Velocity Differential. \textit{Input:} a continuously-updated comparison of clinical-trial enrollment pace and regulatory-review timelines for a given modality across jurisdictions; and regulatory-reform signals (capital-markets reforms, IPO-eligibility changes, clinical-regulatory rule changes) that function as leading triggers for when a velocity advantage is about to become actionable. \textit{Decision logic:} where the data confirms a verified velocity advantage and a reform signal confirms that advantage is investable now, shorten the assumed time-to-value in the valuation model (raising the value attributed to the asset's embedded optionality) and size the position above prevailing consensus, initiated earlier than peer positioning. \textit{Output:} a higher assigned valuation for the asset's optionality, and an earlier, larger position than peer consensus which is bounded by the liquidity limit in Dimension 3.

\textbf{Dimension 3:} Vehicle-Liquidity Structuring. \textit{Input:} position-level liquidity data for each holding (trading volume, free float, estimated days-to-liquidate at various trade sizes) and modeled redemption-stress scenarios for the vehicle as a whole, motivated by the documented tendency of open-end funds holding illiquid positions to face self-reinforcing outflows and to transact at fire-sale prices under stress \citep{Chen2010, Coval2007}. \textit{Decision logic:} apply a fixed ceiling to any single holding's share of the portfolio and maintain a minimum liquidity reserve across the vehicle, sized against a modeled redemption-stress scenario rather than a static compliance rule. \textit{Output:} a hard portfolio-construction constraint (in practice, no single holding exceeding approximately 10\% of the portfolio, and a liquidity reserve maintained at approximately 20\%) that permits the earlier, heavier positions Dimensions 1 and 2 call for to be held without exposing the vehicle, or its investors, to forced-liquidation risk. Dimensions 1 and 2 are the source of this method's distinctive judgment; Dimension 3 is the discipline that makes acting on that judgment safe to execute at scale.

\textbf{Case Study:} The three dimensions were first applied to design, launch, and manage China’s first dedicated cross-border biotechnology fund (CSRC fund code 001984), with the author serving as sole portfolio manager from its February 2019 inception. The strategy was catalyzed by the HKEX Chapter 18A reform in April 2018 \citep{HKEX2018}, which for the first time admitted pre-revenue biotechnology issuers to the Main Board and thereby opened an entire asset class to public capital overnight while it remained largely unpriced by the market. Moving rapidly from regulatory trigger to internal approval and structural design, the vehicle launched in approximately ten months—preceding the asset class's formal benchmark index by nearly a year, and proving that the fund captured, rather than followed, market recognition. The execution delivered exceptional results across multiple dimensions: within sixteen months, the fund returned 127.17\% against a 50.67\% benchmark, while scaling 93-fold as a share of its platform's cross-border AUM. Through the 2020 market-stress period, it ranked first among 276 comparable peer funds nationally (while the peer average turned negative), a resilience directly enabled by Dimension 3's liquidity discipline. A contemporaneous illustration of Dimension 1 is found in Innovent Biologics: its founder's trajectory from a remote village in Zhejiang, through a CAS doctorate and U.S. postdoc training, to executive roles before founding the company in 2011 which exemplifies the multi-factor human capital signal (internationally validated science paired with execution capability) that conventional financial models routinely miss. Crucially, repeatability was proven when the same method, applied concurrently to a structurally distinct domestic-market fund from August 2019, produced comparable top-tier outperformance, confirming that the framework is a systematic, scalable methodology rather than a one-time product of favorable timing.

\subsection{A Multi-Agent AI Framework for
Valuing Clinical-Stage, Cross-Border Biotechnology}

This section describes the proposed framework at the level of architecture and method. Consistent with the confidentiality boundary set out previously, it omits proprietary implementation details such as probability calibrations, and weighting parameters. 

\subsection{Why Multi-Agent Framework?}

Before detailing the design, a foundational question must be addressed: why decompose the task into multiple agents rather than prompt a single, powerful model with a comprehensive briefing. 

\textbf{First, heterogeneous reasoning demands specialization.} Valuing a clinical-stage company is not a uniform cognitive task, but a collection of distinct disciplines performed by different experts in a real firm: parsing trial data like a clinician, evaluating a regulatory pathway like a regulatory affairs specialist, constructing a risk-adjusted net present value (rNPV) like a valuation analyst, and reconciling market-specific pricing like a cross-market trader. These modes draw on disparate knowledge bases and exhibit distinct failure modes. A monolithic model tasked with all four tends to address each superficially, entangling its reasoning, making it impossible to inspect \textit{scientific judgment} in isolation from \textit{valuation arithmetic}. Specialized agents allow each mode of reasoning to be developed deeply, contextualized by domain-specific prompts and \textit{verified independently}, consistent with evidence that role specialization and explicit intermediate reasoning improve performance on compound tasks \citep{Hong2024, Wu2024, Wei2022}: a clinical expert can audit the scientific agent's assessment without wading through valuation math.

\textbf{Second, the architecture must preserve disagreement, rather than average it away.} This is the most critical rationale, linking directly to the framework's core contribution. In this domain, the most valuable signal is frequently the \textit{tension} between conflicting views such as a robust scientific read colliding with a cautious regulatory stance. A single model prompted for a single output tends to resolve such tension internally and silently, smoothing it into a hedged median; the disagreement is precisely the signal a portfolio manager most needs to see but vanishes into the model's hidden states. A multi-agent architecture does the reverse: it forces each perspective to be formulated and stated independently, exposing conflicts as first-class, inspectable objects; multiagent debate has been shown to improve factuality precisely because independently instantiated positions are surfaced and contested rather than silently reconciled \citep{Du2024}. Only once a disagreement is visible can it be weighted domain-specifically. The multi-agent structure is therefore a structural precondition for the opinion-fusion mechanism detailed in following; conflicting expert views cannot be intelligently fused if a monolithic model has already averaged them out of existence.

\textbf{Third, auditability and accountability require decomposition.} Because this framework is designed to support, not replace, a human PM, its outputs must be auditable. A monolithic model emitting a valuation functions as a black box; when it fails, localizing the point of failure is nearly impossible. This concern is sharpened in high-stakes settings, where post hoc explanation of an opaque model is a weaker guarantee than an inherently inspectable decision structure \citep{Rudin2019}, and by the propensity of language models to produce fluent but unsupported assertions \citep{Ji2023}. A layered, multi-agent structure yields decisions with a traceable provenance: a scientific read feeding a success probability, feeding a valuation, adjusted by a cross-market view, and synthesized through a weighted fusion step. When a human PM disagrees, or when an outcome proves the system wrong, the error can be isolated down to a specific agent and corrected. Accountability demands an inspectable chain of reasoning, which architectural decomposition provides.

Taken together, these three reasons establish that the architecture is functionally necessary rather than stylistic: the task is heterogeneous (demanding specialization), its premier signal is conflict (demanding preservation), and it must remain accountable to human oversight (demanding auditability). A monolithic model fails all three criteria. 

\subsection{Framework}
\label{sec:framework}

The framework is organized into the following layers, as illustrated in Figure~\ref{fig:framework_architecture}. The roster is illustrative of the design rather than an exhaustive specification.

\usetikzlibrary{shapes,arrows.meta,positioning,calc}

\begin{figure*}[htbp]
\centering
\begin{tikzpicture}[
    node distance=0.6cm,
    every node/.style={font=\small},
    % 基础节点样式
    box/.style={rectangle, draw=black, thick, rounded corners=3pt, minimum width=11cm, minimum height=1cm, align=center, fill=white},
    % 原创贡献节点样式（带浅蓝色背景，凸显 [★] 贡献）
    contrib/.style={rectangle, draw=blue!50!black, thick, rounded corners=3pt, minimum width=11cm, minimum height=1.1cm, align=center, fill=blue!3!white},
    % 箭头样式
    arrow/.style={-{Stealth[length=2mm, width=2mm]}, thick, black!80},
    % 反馈回路样式
    feedback/.style={-{Stealth[length=2mm, width=2mm]}, thick, dashed, red!60!black}
]
    % 1. Inputs
    \node (inputs) [box, fill=gray!10, minimum height=0.8cm] {
        \textbf{INPUTS:} trial data $\cdot$ mechanism $\cdot$ regulatory path $\cdot$ competitive landscape $\cdot$ multi-market prices
    };
    
    % 2. Analyst Layer
    \node (analyst) [contrib, below=of inputs] {
        \textbf{ANALYST LAYER} \textit{(Parallel Execution)}\\
        \textbf{[$\star$] Scientific \& Clinical Agents} \quad $\vert$ \quad \textbf{[$\star$] Cross-Market Agents (A-share $\cdot$ HK $\cdot$ U.S.)}
    };
    
    % 3. Valuation Agent
    \node (valuation) [contrib, below=of analyst] {
        \textbf{[$\star$] VALUATION AGENT} \textit{(Event-Driven Rebuild)}\\
        rNPV + scenario analysis $\cdot$ probability-weighted binary gates $\rightarrow$ \textbf{Value Range}
    };
    
    % 4. Risk / Portfolio Agent
    \node (risk) [box, below=of valuation] {
        \textbf{RISK / PORTFOLIO AGENT} \textit{(Inherited Shape)}\\
        Binary-event position sizing $\cdot$ Cross-holding correlation modeling
    };
    
    % 5. PM Synthesizer Agent
    \node (synth) [contrib, below=of risk] {
        \textbf{[$\star$] PM SYNTHESIZER AGENT} \textit{(Decision Layer)}\\
        Conflict-type-aware opinion fusion (\S\,\ref{sec:framework}) $\rightarrow$ Allocation + Confidence + Audit Trail
    };
    
    % 6. Human PM
    \node (human) [box, below=of synth, fill=gray!5] {
        \textbf{HUMAN PM} \textit{(Retains Fiduciary Judgment)}\\
        Receives structured, auditable recommendation $\cdot$ Human-in-the-loop override
    };

    % 主干数据流箭头
    \draw [arrow] (inputs.south) -- (analyst.north);
    \draw [arrow] (analyst.south) -- (valuation.north);
    \draw [arrow] (valuation.south) -- (risk.north);
    \draw [arrow] (risk.south) -- (synth.north);
    \draw [arrow] (synth.south) -- (human.north);

    % 反馈回路 (Feedback Loop)
    \draw [feedback] (synth.east) -- ++(1.0,0) node[right, font=\footnotesize, text=red!60!black, align=left] {Unresolved conflicts:\\Iterate re-analysis} |- (analyst.east);

\end{tikzpicture}
\caption{Proposed multi-agent framework architecture and data flow. Boxes marked with $[\star]$ indicate original architectural contributions absent in general financial multi-agent systems.}
\label{fig:framework_architecture}
\end{figure*}
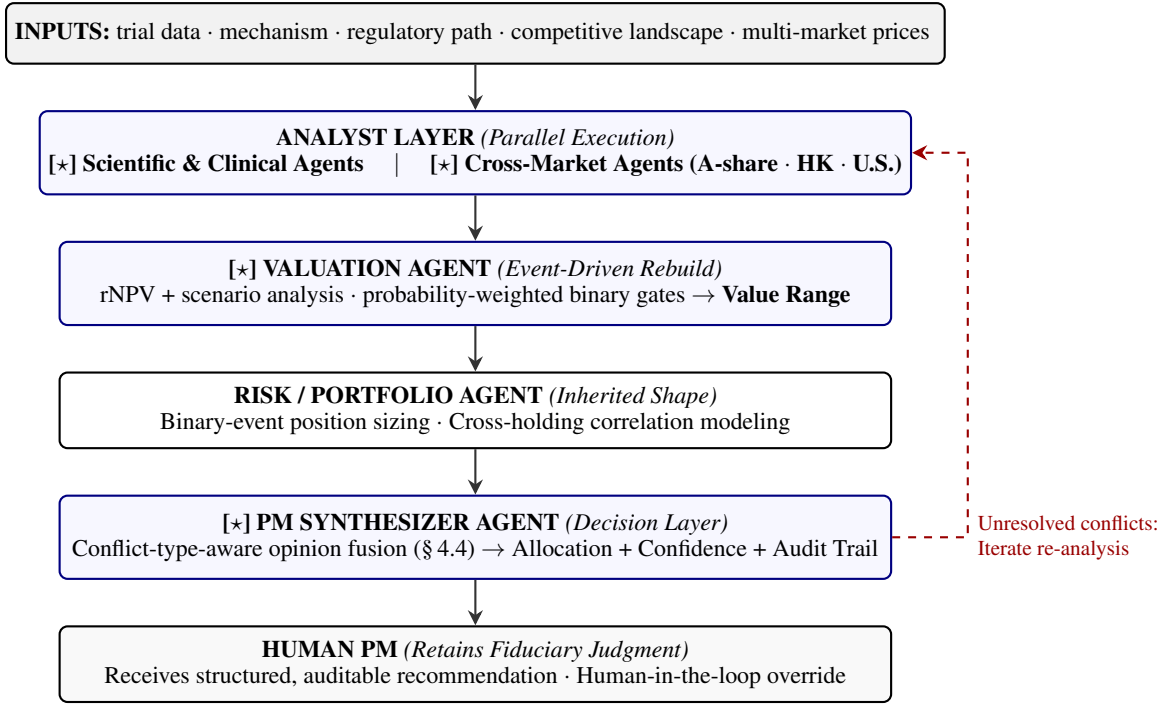

\textbf{Scientific and Clinical Analyst Agents.} This layer has no counterpart in general-purpose financial systems. These agents read and interpret the material a translational scientist or clinical reviewer would evaluate: mechanisms of action, trial designs and endpoints, published and conference-disclosed data, the competitive and standard-of-care landscape, and regulatory pathways. Their output is not a simplistic buy/sell signal, but a structured scientific read. Crucially, each read is accompanied by an explicit rationale and a confidence level, allowing downstream agents (and human auditors) to inspect not only \textit{what} the agent concluded, but \textit{why}. This layer automates the founder-and-science evaluation that Dimension 1 of the human practice performed manually through prospectuses and disclosures.

\textbf{Valuation Agent (Rebuilt).} Bypassing traditional DCF or multiple-based models \citep{Damodaran2012}, this agent operates on event-driven logic: it ingests the scientific analysts' structured reads to construct a risk-adjusted, scenario-based valuation, probability-weighting binary outcomes (trial success/failure, approval/rejection) across developmental stages. The underlying arithmetic follows the risk-adjusted net present value and real-option traditions developed for pharmaceutical assets \citep{Stewart2001, Villiger2005, Kellogg2000, Schwartz2004}, with development costs and timelines calibrated against published industry estimates \citep{DiMasi2016}; the framework's contribution is not this arithmetic but the agentic pipeline that supplies its scientific inputs. It is explicitly designed to generate a defensible valuation for pre-revenue assets with zero earnings, outputting a \textit{range} rather than a point estimate to reflect underlying scientific uncertainty.

\textbf{Cross-Market Agents.} Operating one agent per market context (e.g., A-share, Hong Kong, U.S.), this layer incorporates each market’s distinct valuation conventions, liquidity and currency characteristics, and regulatory rules. Their role is to express the same underlying asset in terms of each individual market and to surface pricing divergences where cross-market friction often generates alpha rather than mere noise. That such divergences persist between economically equivalent claims, and are driven by investor-base segmentation, short-sale constraints, and heterogeneous beliefs rather than by fundamentals, is well established empirically \citep{Froot1999, Mei2009, Karolyi2006}.

\textbf{Risk and Portfolio-Construction Agent.} Because returns in this asset class are driven by binary events, position sizing and diversification obey a logic distinct from cash-flow equities: the portfolio must be constructed to absorb inevitable individual failures without systemic collapse, balancing concentration against the asymmetric payoffs of clinical success in the spirit of growth-optimal sizing under uncertain, repeated binary bets \citep{Kelly1956}. Single-binary failures that would be catastrophic at high concentration must remain survivable. Furthermore, the correlation structure among clinical-stage holdings such as shared mechanisms, indications, or regulators must be explicitly modeled, as these assets can fail concurrently in ways cash-flow equities rarely do. This layer encodes the discipline of Dimension 3, enforcing the liquidity and concentration constraints necessary to hold high-conviction positions safely at scale and to avoid the fire-sale and redemption dynamics that penalize illiquid concentrated books under stress \citep{Chen2010, Coval2007}.

\textbf{PM Synthesizer Agent.} Serving as the core decision layer, this agent integrates the scientific reads, rebuilt valuations, cross-market views, and risk constraints into a final allocation judgment. 

The synthesizer does not merely tally or debate. It applies a domain-specific procedure for reconciling the characteristic conflicts of biotechnology investing, mostly the tension between a favorable scientific read and an unfavorable regulatory or market read. At the architectural level, this procedure:
(a) represents each expert view with an explicit confidence level and rationale rather than a bare numerical score;
(b) weights conflicting views according to the \textit{type} of disagreement; and
(c) propagates residual uncertainty directly into the valuation rather than collapsing it prematurely into a point estimate.
The specific weighting logic reflects specialized investment judgment; the core claim is the architectural principle that opinion fusion in this domain must be conflict-type-aware rather than domain-neutral.

To ground this typology without disclosing calibrations: a conflict between a strong \textit{mechanism} read and a weak \textit{approval-path} read differs fundamentally from a conflict between strong \textit{efficacy} data and a stretched \textit{valuation}. The former addresses scientific and regulatory risk, which belongs inside the probability estimate; the latter addresses pricing and margin of safety. Collapsing both into a single averaged "score," as a domain-neutral system would, discards precisely the nuance a portfolio manager relies upon to decide. The framework's fusion step is thus organized around a taxonomy of conflict types, routing each to the specific layer of the valuation where it belongs.

Crucially, fusion is not restricted to a single forward pass. Certain conflicts can be resolved using already-gathered evidence; others cannot, because the disagreement hinges on a question the analyst layer has not yet examined in sufficient depth such as a regulatory caution tied to a specific precedent, or a mechanism dispute resolvable by a closer inspection of trial data. In such cases, the synthesizer performs the exact function of a human PM when two experts disagree: it sends the question back. The framework therefore incorporates a feedback loop (Figure~\ref{fig:framework_architecture}) through which the synthesizer returns unresolved conflicts to the relevant analyst agents for deeper re-analysis, iterating until the conflict is either resolved or, if irreducible, carried forward into the valuation as widened uncertainty. Information flow is thus bidirectional and iterative rather than a linear pipeline. It’s a property mirroring real investment deliberation that generic single-pass aggregation lacks.

The framework is designed to augment a human PM rather than replace her. Its primary value lies in scale and speed for reading broader literature, covering more companies, surfacing disagreements early, and mitigating cognitive biases by forcing each perspective to be stated with its explicit rationale. Ultimate fiduciary judgment remains human. Consequently, the system industrializes the analytically bound inputs to the PM's decision while leaving the decision itself, and its accountability, firmly in human hands. This clarifies precisely what is being scaled: not the decision, but the bandwidth-constrained analytical work that precedes it.

\subsection{A Synthetic and Anonymized Application Scenario}

The following is a \textit{synthetic, anonymized} scenario constructed from publicly known industry patterns, intended solely to make the framework's operation concrete. \textbf{It does not constitute a performance claim, a backtest, or a description of any actual company, position, or decision.}

Consider a hypothetical clinical-stage firm, "Company X," whose lead asset is a Phase II candidate in an oncology indication, dual-listed in Hong Kong (under Chapter 18A) and mainland-accessible via a cross-border channel. The company has zero revenue. Conventional valuation agents, when tasked with evaluating X, would either return null outputs (due to a lack of earnings to discount) or force an inappropriate valuation multiple. The proposed framework proceeds distinctively:

1. \textbf{Scientific and clinical agents} ingest disclosed Phase I data, the mechanism of action, the Phase II trial design and primary endpoints, and the competitive landscape. They output a structured assessment detailing mechanism plausibility, trial robustness, and estimated success probabilities across remaining developmental gates, each backed by an explicit rationale.
2. \textbf{The valuation agent} translates this read into a risk-adjusted net present value (rNPV) in the standard pharmaceutical formulation \citep{Stewart2001, Villiger2005}: it models potential payoffs upon approval, probability-weights binary milestones across developmental stages, and nets out costs and timelines, yielding a defensible value range rather than a spurious point estimate.
3. \textbf{Cross-market agents} express this valuation in terms of Hong Kong and mainland-channel contexts, identifying structural divergences in liquidity, investor bases, and regulatory nuances. They reconcile these factors into a unified perspective while keeping pricing spreads explicit.
4. \textbf{The PM synthesizer} encounters a hallmark domain conflict: a constructive scientific read colliding with regulatory approval uncertainties. Rather than resorting to averaging or simple voting, it applies conflict-type-aware weighting, propagates the regulatory ambiguity into a widened valuation range, and generates an allocation recommendation accompanied by explicit confidence levels and a full audit trail.
5. \textbf{The risk and portfolio-construction agent} sizes the position to ensure that a binary failure of Company X remains survivable within the portfolio, accounting for its correlation structure with other clinical-stage holdings.

The objective of this walkthrough is not to present specific numerical outputs, but to illustrate the \textit{shape} of the process: an event-driven valuation capability absent in general financial systems, reconciled across international markets, featuring domain-specific fusion of conflicting expert views, and delivering a structured, auditable recommendation to a human portfolio manager rather than a black-box signal.

\section{Future directions}
\label{sec:future}

The natural next steps are empirical: validating the framework's valuations against realized outcomes on public, disclosed events; extending the approach to adjacent pre-revenue asset classes (for example, medical devices and platform technologies with similar binary-milestone structures); and refining the human-in-the-loop interface so that the system's outputs are maximally auditable and useful to a working PM. The observation underlying Dimension 2 of the human practice that a market-opening regulatory reform is a leading indicator of capital formation into a previously inaccessible asset class has, notably, already recurred beyond biotechnology: a 2023 reform permitting specialist technology companies (including artificial-intelligence companies) to list under a comparable pre-revenue framework \citep{HKEX2023} produced a capital-formation pattern structurally analogous to the one identified for biotechnology in 2018–2019, independent evidence that this component describes a general market mechanism rather than a single-sector observation. Each of these directions is a program of work in its own right; the contribution of this paper is the framework, and the proven human method beneath it, that make them possible.

\textit{Disclaimer: This whitepaper presents a conceptual framework for discussion, grounded in the author's documented prior investment practice. The framework itself does not constitute investment advice, and it makes no representation of realized or expected investment performance for any AI system. Historical results reported for the author's human practice are statements about that prior practice only. Proprietary implementation details are intentionally withheld.}

\newpage
\bibliographystyle{plainnat}
\bibliography{references}

\end{document}